\documentclass[11pt, a4paper]{article}
\pdfoutput=1

\usepackage[height=22cm,width=17cm, centering]{geometry}
\usepackage{setspace}

\usepackage[utf8]{inputenc}

\usepackage{amsmath}
\usepackage{amssymb}
\usepackage{graphicx}
\usepackage{subcaption, bm}
\usepackage{color}
\usepackage[sort&compress, numbers, merge]{natbib}

\definecolor{purple}{rgb}{0.5 ,0, 0.7}
\definecolor{bluegreen}{rgb}{0, 0.45, 0.35}
\definecolor{fig_purple}{rgb}{0.61 ,0.15, 1.}
\definecolor{fig_cyan}{rgb}{0, 0.87, 0.77}

\usepackage{hyperref}
\hypersetup{colorlinks=false, linkcolor=bluegreen, citecolor=purple, urlcolor=blue}

\allowdisplaybreaks

\begin{document}

\begin{titlepage}
\begin{center}
\leavevmode \\
\vspace{ 0cm}

\hfill {\small IPMU 19-0178} \\
\hfill {\small CTPU-PTC-19-36}

\noindent
\vskip 2 cm
 {\LARGE Gauge Independence of Induced Gravitational Waves}

\vglue .6in

{ 
Keisuke Inomata${}^{a, b}$, Takahiro Terada${}^c$
}

\vglue.3in

{\small 
\textit{ ${}^a$  ICRR, University of Tokyo, Kashiwa, 277-8582, Japan  \\
${}^b$ Kavli IPMU (WPI), UTIAS, University of Tokyo, Kashiwa, 277-8583, Japan \\
${}^c$ Center for Theoretical Physics of the Universe, \\ Institute for Basic Science (IBS),  Daejeon, 34126, Korea }
}

\end{center}

\vglue 0.7in

\begin{abstract}
We study gauge (in)dependence of the gravitational waves (GWs) induced from curvature perturbations.  For the GWs produced in a radiation-dominated era, we find that the observable (late-time) GWs in the TT gauge and in the Newtonian gauge are the same in contrast to a claim in the literature.  
We also mention the interpretation of the gauge dependence of the tensor perturbations which appears in the context of the induced GWs.
\end{abstract}

\end{titlepage}

\newpage


\section{Introduction}\label{sec:intro}

One of the main predictions of cosmological inflation is the generation of the cosmic perturbations.  So far, there is only an upper bound on the amplitude of tensor perturbation, but that of scalar perturbations, namely the curvature perturbations, has been well established by the precise observations of the cosmic microwave background and the large-scale structure~\cite{Aghanim:2018eyx, Akrami:2018odb}.  
Although these observations provide us information on the large scale, we cannot access much smaller scales due to the Silk damping of the radiation perturbations or the nonlinear growth of the matter perturbations.  
Such information on the small-scale perturbations is desirable to understand the global picture of the inflaton potential and particle physics behind it.  

Gravitational waves (GWs) are a useful probe of small-scale curvature perturbations~\cite{Alabidi:2012ex, Alabidi:2013wtp, Orlofsky:2016vbd, Inomata:2018epa, Byrnes:2018txb, Ben-Dayan:2019gll} since if the latter is enhanced, a sizable amount of GWs is produced at the second order of the cosmological perturbation. We call it the induced GWs\footnote{
It is also called the second-order GWs, secondary GWs, or scalar-induced GWs, etc. 
} (see early works~\cite{Mollerach:2003nq, Ananda:2006af, Baumann:2007zm, Assadullahi:2009nf} and recent developments~\cite{Espinosa:2018eve, Kohri:2018awv, Cai:2018dig,  Bartolo:2018rku, Unal:2018yaa, Inomata:2019zqy, Inomata:2019ivs, Yuan:2019udt, Bartolo:2019oiq, Yuan:2019wwo}).  
 Such an enhancement of the curvature perturbations are often considered in the context of the primordial black hole (PBH)~\cite{Saito:2008jc, Saito:2009jt, Bugaev:2009zh} (see also Refs.~\cite{Clesse:2018ogk, Byrnes:2018txb, Wang:2019kaf,  Cai:2019elf, Bartolo:2019zvb, Chen:2019xse, Hajkarim:2019nbx}).  
 PBHs have recently been studied by many authors mainly because they can be a dark matter (DM) candidate and may explain the binary black hole merger events observed by LIGO/Virgo~\cite{LIGOScientific:2018mvr}.  
If the PBHs explain the DM or LIGO/Virgo events, the induced GWs can be observed by the future projects, such as LISA~\cite{Audley:2017drz} and SKA~\cite{Janssen:2014dka}, and even if there are few PBHs in the Universe, the future observations of the induced GWs could determine or constrain the amplitude of the small-scale perturbations.

In contrast to the linear perturbation theory, the second-order (induced) tensor perturbations are known to have gauge dependence. 
A large change of the power spectrum of the tensor perturbations induced in a matter-dominated (MD) era was reported in Ref.~\cite{Hwang:2017oxa}.  
More aspects of the gauge dependence were studied in Refs.~\cite{Matarrese:1997ay, Arroja:2009sh, Domenech:2017ems, Gong:2019mui, Tomikawa:2019tvi}.  

In the context of the tensor perturbations induced by scaler perturbations, the tensor perturbation can be divided into two parts. 
One is the freely propagating tensor perturbation, following the equation of motion without any source.
This kind of tensor perturbations is widely known as \emph{gravitational waves} and the time dependence can be written as $h_{ij} \propto \sin(k\eta)/a$ or $\cos(k\eta)/a$ ($a$: scale factor), as is well known.
Although they couple with scalar perturbations at their production, they finally decouple from the scalar perturbations and freely propagate.
Once they decouple from the scalar perturbation, they are independent of the scalar perturbations and therefore do not depend on the gauge.
The other is the tensor perturbation coupling with the scalar perturbations, in which the freely propagating tensor perturbations are included only until they decouple from the scalar perturbations.
Since this kind of tensor perturbation is controlled by the scalar perturbations, the time dependence of this tensor perturbation inherits those of the scalar perturbations.
Note that the gauge dependence appears in this kind of tensor perturbation.
This tensor perturbation is also often called a gravitational wave in many references, but in this paper, to discriminate these two kinds of tensor perturbations, we call only the freely propagating tensor perturbation a \emph{gravitational wave}.

Recently, it was claimed\footnote{
One of the main claims in Ref.~\cite{DeLuca:2019ufz} is  that the description of the detection of GWs with a space-based interferometer is most straightforward in the TT gauge, and they advocate the use of the TT gauge (if the strength of the induced GWs is gauge dependent at all).  The argument is as follows.
The size of the space-based interferometer is typically not sufficiently large compared to the wavelength of the GWs.
In such a situation, one cannot take the proper detector frame (Fermi normal coordinates) for the whole experimental region. The completely relativistic treatment is required, and they give at least two reasons that motivate one to use the TT gauge.
For one thing, one has to otherwise add the integral of the gravitational potential along the photon path to make the time shift (measured as the phase shift of the laser) gauge invariant in other gauges.  Note that the result of the TT gauge reduces to the simplest case where the description in the proper detector frame is possible.  For the other, the sensitivity curve of LISA is calculated in  the TT gauge. 
} in Ref.~\cite{DeLuca:2019ufz} that the power spectrum of the induced GWs (freely propagating tensor perturbations) calculated in the transverse-traceless (TT) gauge (also known as synchronous gauge) is reduced by at least one order of magnitude compared to that calculated in the Newtonian gauge (also called the Poisson gauge, the longitudinal gauge, or the zero-shear gauge), commonly used in the literature for the GWs produced during a radiation-dominated (RD) era.

However, we do not expect that the induced GWs depend on the gauge as we mentioned above.
The gauge independence of the induced GWs can also be expected from the coincidence of the GWs calculated in the Newtonian gauge and the uniform curvature gauge (also called the flat gauge), shown in Ref.~\cite{Tomikawa:2019tvi} (see the case of $w>0$ in the reference).
In the present paper, we revisit the calculation in Ref.~\cite{DeLuca:2019ufz} and find that the difference between the tensor perturbations induced during a RD era in the TT gauge and in the Newtonian gauge decreases on subhorizon scales, and therefore the late-time (i.e.~observable) GWs are the same in both gauges, in contrast to Ref.~\cite{DeLuca:2019ufz}.  
We also mention the interpretation of the gauge dependence reported in Ref.~\cite{Tomikawa:2019tvi} and an important role played by the diffusion damping effect at the end of this paper.

\section{Induced Gravitational Waves in TT gauge}\label{sec:gw}
In this section, we calculate the induced GWs in the TT gauge and compare them with those in the Newtonian gauge known in the literature.
We take the convention in Refs.~\cite{Inomata:2016rbd, Kohri:2018awv}.  
The comparison of the conventions in the literature is given in Appendix~\ref{sec:convention}.
To avoid the repetition of steps of derivation, we write only the essential equations for a review.
Details of derivation can be found in Refs.~\cite{Inomata:2016rbd, Kohri:2018awv}  (see also Ref.~\cite{Matarrese:1997ay} for early works on the formulation in the TT gauge).

The GW energy density is given by~\cite{Maggiore:1999vm}
\begin{align}
\rho_\text{GW} = &  \frac{1}{16 a^2} \overline{\langle h'_{ij} h'_{ij} \rangle} ,
\end{align}
where $h_{ij}$ is the freely propagating tensor perturbation, and the overline denotes the oscillation average.
Thanks to the oscillation average, we may calculate the spatial derivatives $h_{ij,k}h_{ij,k}$ instead of the time derivatives, which is justified on subhorizon scales.

The current value of the power spectrum of the induced GWs per a logarithmic wavenumber bin $\Omega_\text{GW} (\eta, k) \equiv \rho_\text{GW}(\eta, k) / \rho_\text{total} $ with $\rho_\text{GW}(\eta) = \int \text{d}\ln k \,  \rho_\text{GW}(\eta, k)$ is obtained from 
\begin{align}
\Omega_\text{GW}(\eta_0 , k) = 0.387 \, \Omega_\text{r} \left( \frac{g_* (T_\text{c})}{106.75} \right) \left( \frac{106.75}{g_{*,s}(T_\text{c})}\right)^{\frac{4}{3}} \Omega_\text{GW} (\eta_\text{c}, k),   \label{Omega_GW_now}
\end{align}
where $\eta = \eta_0$ is the current conformal time; a subscript c denotes a time at which the tensor perturbation is dominated by the freely propagating tensor perturbation\footnote{If we define the parameter $\Omega_\text{GW}$ as the energy density parameter of the total tensor perturbations, as in many references, the subscript c denotes a time at which $\Omega_\text{GW}$ reaches the asymptotic constant value. }; $g_{*}$ and $g_{*,s}$ are the numbers of relativistic degrees of freedom for the energy and entropy density, respectively; and $\Omega_\text{r}$ is the current energy density fraction of radiation. 
This formula assumes the production of the GWs in a RD era.  The constant value of $\Omega_\text{GW}$ can be calculated as
\begin{align}
\Omega_\text{GW}(\eta_\text{c}, k) = \frac{1}{6} \left( \frac{k}{\mathcal{H}} \right)^2 \int_0^\infty \text{d} v \int_{|1-v|}^{1+v} \text{d}u \left( \frac{4 v^2 - (1+ v^2 - u^2 )^2}{4 u v} \right)^2 \overline{I^2(u,v,x_\text{c})} \mathcal{P}_\zeta (u k) \mathcal{P}_\zeta (v k) ,  \label{Omega_GW_constant}
\end{align}
where $x \equiv k \eta$ is a dimensionless time, $\mathcal{H}= (\text{d}a/\text{d}\eta) / a$ is the conformal Hubble parameter, $\mathcal{P}_\zeta (k)$ is the power spectrum of the primordial curvature perturbations, and the late time limit $x \to \infty$ is practically used for evaluation at $x_\text{c} = k \eta_\text{c}$.
The function $I(u,v,x)$ is defined as
\begin{align}
I(u,v,x) = \int_0^x \text{d} \bar{x} \frac{a(\bar{\eta})}{a(\eta)} k G_k (\eta , \bar{\eta}) f(u,v,\bar{x}),  \label{I_definition}
\end{align} 
where $\bar{x} = k \bar{\eta}$, and  $G_k(\eta,\bar{\eta})$ is the Green function for the tensor perturbations, which is given as $k G_k (\eta, \bar{\eta}) = \sin (x - \bar{x})$ in the production in a RD era.
In this expression, $u$ and $v$ are symmetric under their exchange.\footnote{  \label{fn:symmetry}
The power spectrum of the induced GWs is nothing but a 2-point correlation function, which can be written as a 4-point function of scalar source modes.
In the absence of non-Gaussianity, the 4-point function can be expressed in terms of two different combinations of 2-point functions (two ways of Wick contraction).  From this and from the $(u \leftrightarrow v)$ symmetry of the rest of the integrand as well as the integration region, the remaining asymmetry under the exchange of $u$ and $v$, if any, must vanish. Once this is understood, one can freely symmetrize $f(u,v,x)$ or equivalently $I(u,v,x)$ as a convention.
} 

Now, the final missing piece in the above equation is the explicit definition of the source function $f(u,v,\bar{x})$, which is different between the TT gauge and the Newtonian (or some other) gauge. The difference arises only from this.
It can be derived from the expression of the scalar-scalar source term in the equation of motion for the second-order tensor perturbation, Eq.~(46) of Ref.~\cite{Gong:2019mui}.  In terms of the transfer functions $T_\psi$ and $T_\sigma$ of the scalar perturbation variable $\psi$ and the shear potential $\sigma$ in the TT gauge, it is given by
\begin{align}
f(u,v,x) \equiv & - \left( \frac{2}{3}\right)^2 \left(  - T_\psi (u x) T_\psi (v x) + \frac{3 \sqrt{3} v}{2 u} T'_\psi (v x) T_\sigma (u x) + \frac{3 \sqrt{3} u}{2 v} T'_\psi (u x) T_\sigma (v x)  \nonumber \right. \\
& \left.  \qquad \qquad \quad + \frac{3}{2 u v} T_\sigma (u x) T_\sigma (v x)  - u v x^2 T'_\psi (u x) T'_\psi (v x) \right) ,  \label{fTT}
\end{align}
\begin{align}
T_\psi (x) = & \frac{9-9\cos(x/\sqrt{3})}{x^2}, &
T_\sigma (x) = & \frac{-3\sqrt{3} + 9 \sin (x/\sqrt{3})}{x^2},
\end{align}
where the prime denotes the differentiation with respect to the argument, e.g. $T'(ux) \equiv \frac{\text{d} T(ux)}{\text{d} (ux)}$.  
More explanations on the transfer functions and the choice of an integration constant are given in Appendix~\ref{sec:convention}.
A difference of an overall factor from Ref.~\cite{DeLuca:2019ufz} is just because of different conventions.  We find the coefficient of the first term inside the parenthesis in Eq.~\eqref{fTT} $-1$, but $2$ is used in Ref.~\cite{DeLuca:2019ufz}.  
Also, our definition of the transfer functions is 3 times larger than that in Ref.~\cite{DeLuca:2019ufz} for the transfer function of the gauge invariant Bardeen potential $\Psi$ to be unity for $\eta \to 0$.
Therefore, we expect an amplitude and shape of the power spectrum of the induced GWs different from those of Ref.~\cite{DeLuca:2019ufz}.

Similarly to the case of the Newtonian gauge studied in Refs.~\cite{Espinosa:2018eve, Kohri:2018awv}, we can analytically integrate Eq.~\eqref{I_definition} using the method of Ref.~\cite{Ananda:2006af}. The result is long, so it is written in Appendix~\ref{sec:analytic_formulae}.
First of all, it is different from the case of the Newtonian gauge for finite $x = k \eta$.  
The difference between the two gauges is given by
\begin{align}
\mathcal{I}^{\text{(Newtonian gauge)}} (u,v,x) - \mathcal{I}^{\text{(TT gauge)}} (u,v,x) = & \frac{18}{u^2 v^2 x} \left( 1 - \frac{\sqrt{3}}{u x} \sin \frac{u x}{\sqrt{3}}  \right) \left( 1 - \frac{\sqrt{3}}{v x} \sin \frac{v x}{\sqrt{3}}  \right)  \nonumber \\
\simeq & \min \left[ \frac{x^3}{18}, \, x,  \, \frac{18}{u^2 v^2 x}  \right],   \label{coincedence}
\end{align}
 where $\mathcal{I}(u,v,x) \equiv x I (u,v,x)$, and the superscripts denote in which gauge it is evaluated. 
 We introduce this combination $\mathcal{I}$ as its oscillation average asymptotes to a constant while $I$ decreases by redshift.
 In the second equality, we have used the fact that $\max [u^2 , v^2] \simeq 1$ when $\min [u, v] \ll 1$. 
 From the above expression, it is clear that the gauge difference decreases in time on subhorizon scales, and the asymptotic constant values, which are relevant for direct observations, coincide with each other.  That is, $ \lim_{x \to \infty}  \mathcal{I}^{\text{(TT gauge)}} (u,v,x)=  \lim_{x \to \infty}   \mathcal{I}^{\text{(Newtonian gauge)}} (u,v,x)$. 
  Therefore, we can use the same formula~\cite{Kohri:2018awv} (see also Ref.~\cite{Espinosa:2018eve})
\begin{align}
\lim_{x \to \infty} \overline{{\mathcal{I}}^2(u,v,x)} =& \frac{1}{2} \left( \frac{3(u^2+v^2-3)}{4 u^3 v^3 } \right)^2 \left( \left( -4uv+(u^2+v^2-3) \log \left| \frac{3-(u+v)^2}{3-(u-v)^2} \right| \right)^2  \right. \nonumber \\
&  \left. \qquad \qquad \qquad \qquad  \phantom{\left(\left|\frac{(uv)^2}{(uv)^2}\right|\right)^2}  + \pi^2 (u^2+v^2-3)^2 \Theta ( u+v-\sqrt{3}) \right), \label{I_RD_osc_ave}
\end{align}
in both the TT gauge and the Newtonian gauge, where $\Theta ( \cdot)$ is the Heaviside step function.  Thus, once the power spectrum of the primordial curvature perturbations $\mathcal{P}_\zeta (k)$ is specified, one can use Eqs.~\eqref{Omega_GW_now}, \eqref{Omega_GW_constant}, and \eqref{I_RD_osc_ave} to compute the power spectrum of the induced GWs, and the result is irrespective of the choice of the TT gauge or the Newtonian gauge.  Explicitly, 
\begin{align}
\Omega_\text{GW}^\text{(TT gauge)} (\eta_0 , k) = \Omega_\text{GW}^\text{(Newtonian gauge)} (\eta_0 , k),
\end{align}
and an analogous formula with $\eta_0$ replaced by $\eta_\text{c}$ hold.

\begin{figure}[htb!]
 \centering
   \subcaptionbox{not small $u$, $v$ \label{sfig:I_generic}}
{\includegraphics[width=0.49 \columnwidth]{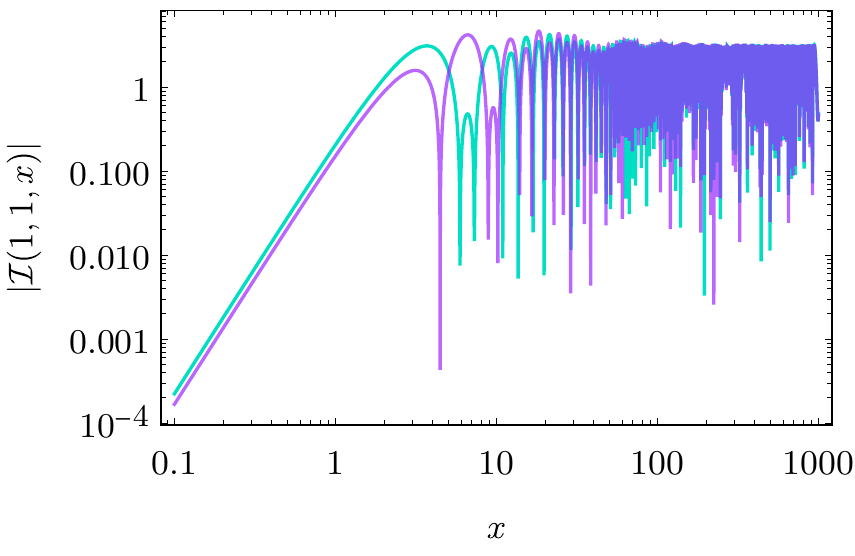}}~
   \subcaptionbox{small $u$ or $v$ \label{sfig:I_small}}
{\includegraphics[width=0.49 \columnwidth]{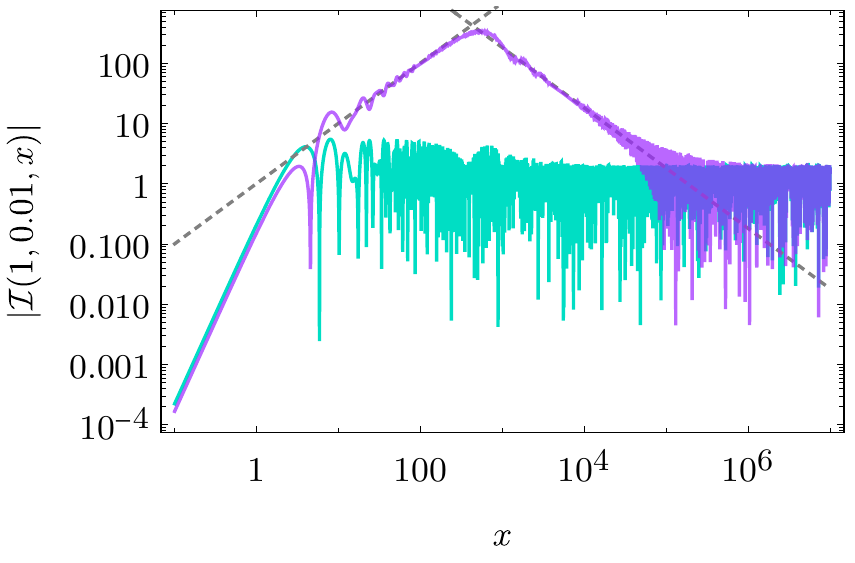}}
  \caption{Comparison of time evolution of the function $\left| \mathcal{I}(u,v,x)\right|$ in the Newtonian gauge (cyan) and in the TT gauge (purple). The values of $u$ and $v$ are set as $(u = 1, v = 1)$ in the left panel, and $(u = 1, v = 0.01)$ in the right panel. The gray dotted lines are $x$ and $18/(u^2 v^2 x)$.}
 \label{fig:time_evolution}
  \end{figure}

In the remainder of this section, let us study the behavior of $\mathcal{I}(u,v,x)$ in each gauge  and their differences for finite $x$.
Examples of the time evolution of $\mathcal{I}(u,v,x) $ for given sets of $u$ and $v$ in both gauges are shown in Fig.~\ref{fig:time_evolution}.  In the figure, $x = k \eta$ can be interpreted as the time parameter in the unit in which the horizon entry of the tensor perturbation ($k=\mathcal{H}$) occurs at $x = 1$.  
The variables $u$ and $v$ are to be interpreted as the ratio of the length scales of the tensor perturbation and the typical or relevant scalar source.  
For example, if we consider the monochromatic power spectrum of the curvature perturbation at $k_*$, taken in Ref.~\cite{DeLuca:2019ufz}, the relevant values of  $u$ and $v$ are $k_* / k$.  

Let us begin with the small $x$ (superhorizon) limit.  The leading order terms of $I(u,v,x)$ in this limit are $(2/9) x^2$ in the Newtonian gauge\footnote{ \label{fn:correction}
It is written as $(1/2)x^2$ in Ref.~\cite{Kohri:2018awv}, but an overall factor was missing.
} and $(1/6) x^2$ in the TT gauge. In terms of $\mathcal{I}$, the difference is thus $(1/18) x^3$.  

After the horizon entry of the scale of the would-be GWs, the difference in the two gauges  becomes large for small $u$ or $v$ (right panel of Fig.~\ref{fig:time_evolution}).  
This should come from the gauge dependence of the part of the tensor perturbations coupling with scalar perturbations.
 After a while, the difference of $\mathcal{I}$ decreases slowly as $\propto x^{-1}$.  
 Finally, the difference becomes smaller than the asymptotic value at a late time $x \gg 1$, which means the tensor perturbations in the TT gauge are dominated by the freely propagating tensor perturbations (GWs).
 Let us see these behaviors quantitatively below.

Typically, the Newtonian gauge results (cyan line) reach the asymptotic value as early as its horizon entry (both in the left and right panel).
TT gauge results (purple line) are similar if $u$ and $v$ are not smaller than 1 (left panel), but this changes if $u$ or $v$ are small, i.e.~$uv \lesssim 1$ (right panel).\footnote{Note that $u$ and $v$ cannot be small simultaneously because of momentum conservation: see the integration region in Eq.~\eqref{Omega_GW_constant}. Also, $uv \ll 1$ is possible only at the corners of the integration region, so the net effect of gauge dependence after integration over $u$ and $v$ is suppressed.}
In this case, $\mathcal{I}$ in the TT gauge once increases as $x$ for $3 \sqrt{2} \lesssim x \lesssim 3\sqrt{2}/(uv)$ and then decreases as $18/(u^2 v^2 x)$ for $x \gtrsim 3\sqrt{2}/(uv)$ until it reaches the asymptotic value, whose $(u, v)$ dependence is not simple [see Eq.~(\ref{I_RD_osc_ave})]. 
The peak of $\mathcal I$, given as $x \simeq 3\sqrt{2}/(uv)$, roughly corresponds to the horizon entry of the scalar perturbation with the wavenumber $\min [ u k, v k]$.
This means that once both the scalar perturbations with their wavenumbers being $u k$ and $v k$ enter the horizon, the gauge dependence of the tensor perturbations starts to decrease and finally becomes negligible.
This behavior is similar to the decay of the gauge dependence for the energy density perturbation on subhorizon scales~\cite{Ma:1995ey}. 
In summary, the gauge dependence vanishes well after the tensor mode and the scalar modes enter the horizon. 
Note again the difference between the tensor perturbations in the two gauges at a finite time is due to the part of those coupling with the scalar perturbations.

\section{Discussion}\label{sec:}

In this paper, we reconsider the results in Ref.~\cite{DeLuca:2019ufz} and show that the late-time (e.g.~observable) GWs calculated in the TT gauge and in the Newtonian gauge coincide with each other,  in contrast to Ref.~\cite{DeLuca:2019ufz}. 
Here, the ``late time'' limit means the subhorizon limit: time much after the horizon entry of the relevant scalar source modes as well as the tensor mode. 
In addition, the calculation in the uniform curvature gauge (flat gauge) coincides with the Newtonian-gauge result too~\cite{Tomikawa:2019tvi}. 
The main reason for the agreement in these gauges is because the tensor perturbations are finally dominated by the freely propagating tensor perturbations (GWs) in all of the three gauges. 

On the other hand, the tensor perturbations could be different in the comoving gauge even in the subhorizon limit during a RD era, according to Ref.~\cite{Tomikawa:2019tvi}. 
We can find that the difference between the Newtonian gauge and the comoving gauge comes from the terms whose oscillations are of the form $\cos \left( (u \pm v) k \eta / \sqrt{3} \right)$ or $\sin \left( (u \pm v) k \eta / \sqrt{3} \right)$, which are not the oscillations of \emph{gravitational waves}, $\cos (k \eta)$ or $\sin (k \eta)$; see Eq.~(65) in Ref.~\cite{Tomikawa:2019tvi}.
This indicates that the gauge dependence appears only in the tensor perturbations coupling with scalar perturbation, not in the gravitational wave.\footnote{
A corollary of this discussion is that the large enhancement of the induced GWs found in Ref.~\cite{Inomata:2019ivs}, which one may suspect a gauge artifact, is not a gauge artifact. They are produced in a RD era, just after a sudden reheating transition, and the late-time time dependence of $I(u,v,x)$ is that of genuine gravitational waves, i.e.~$\cos (k \eta)/a$ or $\sin (k \eta)/a$.
}

We also comment on the gauge dependence from the viewpoint of the diffusion damping~\cite{Silk:1967kq}, which occurs on small scales, since many authors focus on the GWs induced by the small-scale scalar perturbations (often in the context of PBHs).
After inducing GWs, the small-scale perturbations finally decay exponentially, $\propto \exp (- k^2 / k_\text{D}^2 (t))$, within the diffusion scale $k_\text{D}^{-1}(t)$ because the free-streaming length of some species, such as neutrinos or photons, increases as the temperature decreases (see e.g.~Ref.~\cite{Jeong:2014gna} for the time dependence of the damping scale).
This means that the scalar perturbations inducing the GWs during the RD era get exponentially suppressed in most cases, and therefore the gauge dependence of the tensor perturbations would disappear. 
Although we leave the analysis of the diffusion damping effect on the evolutions of the tensor perturbations for future work, we expect that the gauge dependence that appears in the comoving gauge~\cite{Tomikawa:2019tvi} would disappear after the exponential damping of the scalar perturbations.
This is because if the scalar perturbations are exponentially suppressed and can be negligible, the difference between the tensor perturbations in two different gauges, which should be written in terms of the scalar perturbations, can also be negligible. 
In summary, gauge dependence disappears due to the diffusion damping, so the late-time GWs should be unique (the same as the one calculated in the Newtonian gauge).

Note that the situation is different for the tensor perturbation induced by the scalar perturbations during the late MD era.
In this case, the scalar perturbations continue to couple with the tensor perturbations on subhorizon scales even in the Newtonian gauge because of the growing matter perturbations. Therefore, the tensor perturbations are dominated by the ones coupling with the scalar perturbations. That is, this kind of tensor perturbations can be larger than the GWs induced during the RD era on large scales~\cite{Baumann:2007zm}.
This is why the tensor perturbation easily depends on the gauge during a MD era. 
For example, in Appendix~\ref{subsec:md_era_i_formula}, we show that the function $I$ in the TT gauge is different from that in the Newtonian gauge even at late time.
However, we should notice that this kind of tensor perturbation is not a \emph{gravitational wave}. 
The tensor perturbation is time-independent in the Newtonian gauge and does not behave as radiation~\cite{Baumann:2007zm,Assadullahi:2009nf,Kohri:2018awv}.
Since most observations assume that the tensor perturbations are \emph{gravitational waves}, the observational sensitivities for this kind of tensor perturbation will be different from those for conventional gravitational waves.
Also, the observation of this tensor perturbation might require a discussion about the suitable gauge for the observation because of its gauge dependence.

\section*{Note added in proof}
The results of the latest version of Refs.~\cite{DeLuca:2019ufz, Yuan:2019fwv} agree with ours.

\section*{Acknowledgments}
This work was supported in part by World Premier
International Research Center Initiative (WPI Initiative), MEXT,
Japan, the JSPS Research Fellowship for Young Scientists (KI),
JSPS KAKENHI Grants No.~JP18J12728 (KI), Advanced Leading Graduate Course for Photon Science (KI), and IBS under the project code, IBS-R018-D1 (TT).
We used the Mathematica package xPand~\cite{Pitrou:2013hga} during our work.

\appendix
\section{Convention}\label{sec:convention}
We use the reduced Planck unit, $c = \hbar = 8 \pi G = 1$, and basically follow the notations and conventions in Refs.~\cite{Inomata:2016rbd, Kohri:2018awv}.  

The line element is parametrized as 
\begin{align}
\text{d}s^2 = - a^2  (1 + 2 \phi ) \text{d} \eta^2 + 2 a^2 B_{,i} \text{d} \eta \text{d} x^i + a^2 \left(  (1 - 2 \psi) \delta_{ij} + 2 E_{,ij} + \frac{1}{2} h_{ij} \right) \text{d}x^i \text{d}x^j,
\end{align}
where we only write the relevant perturbation variables. The quantities $\phi$, $\psi$, $B$ and $E$ are first-order scalar perturbations, and $h_{ij}$ is the second-order transverse traceless tensor perturbation.   
 The comma denotes differentiation: for example, $E_{,ij} = \partial_i \partial_j E$.  The TT gauge is defined as the gauge where $\phi  = B = 0$.  On the other hand, the Newtonian gauge is defined as $E = B = 0$. 

The normalizations and signs of $\psi$, $E$, and the shear potential $\sigma = \partial E / \partial \eta$ in the TT gauge are the same as those in Ref.~\cite{DeLuca:2019ufz}.  If we pick up the second-order component of $h_{ij}$ [see their Eq.~(A.12g)], it is also the same in the coordinate space.  The convention of the Fourier transform is different, but it does not affect most parts of the calculation.
When there is no anisotropic stress, the following equations are satisfied in each gauge:
\begin{align}
\phi =& \psi \qquad (\text{Newtonian gauge}), \\
\frac{\partial \sigma}{\partial \eta} =& -(\psi + 2 \mathcal H \sigma) \qquad (\text{TT gauge}).
\end{align}

The relation of perturbation variables in the TT gauge and in the Newtonian gauge is given by Eqs.~(5.2) and (5.3) in Ref.~\cite{DeLuca:2019ufz}:
\begin{align}
\psi^{\text{(TT gauge)}} (\eta, \bm{k}) =& \Psi  + \mathcal{H}(\eta) \left(  \int_0^\eta  \frac{a(\bar{\eta})}{a(\eta)} \Psi  (\bar{\eta}, \bm{k}) \text{d} \bar{\eta}  -  \frac{\mathcal{D}_1 (\bm{k})}{a(\eta)} \right),   \\
\sigma^{\text{(TT gauge)}} (\eta, \bm{k}) =& - \int_0^\eta  \frac{a(\bar{\eta})}{a(\eta)} \Psi  (\bar{\eta}, \bm{k}) \text{d} \bar{\eta}  +  \frac{\mathcal{D}_1 (\bm{k})}{a(\eta)},
\end{align}
where $\Psi$ is a gauge invariant Bardeen potential which reduces to $\psi$ in the Newtonian gauge, and $\mathcal{D}_1 (\bm{k})$ is an integration constant which does not depend on $\eta$.  
We take $\mathcal{D}_1 (\bm{k}) = 0$, which corresponds to $\mathcal{C} (\bm{k}) = -2\zeta(\bm{k})/k^2$ in Ref.~\cite{DeLuca:2019ufz}: for comparison, see also their footnote 6.  
The transfer functions of $\psi = \psi^{\text{(TT gauge)}}$ and $\sigma = \sigma^{\text{(TT gauge)}}$ during a RD era in the TT gauge are defined as~\cite{DeLuca:2019ufz} \footnote{We take the sign notation for the curvature perturbation which is opposite to that taken in Ref.~\cite{DeLuca:2019ufz}.} 
\begin{align}
\label{eq:psi_zeta_rel}
\psi (\eta, \bm{k}) = & -\frac{2}{3} \zeta (\bm{k}) T_\psi (\eta , k) , \\
\label{eq:sigma_zeta_rel}
\sigma (\eta, \bm{k} ) =& -\frac{2}{3} \zeta (\bm{k}) \frac{\sqrt{3}}{k} T_\sigma (\eta, k),
\end{align}
where $\zeta(\bm{k})$ is the primordial curvature perturbations.

Table~\ref{table:notation} summarizes the correspondence of the notations in this paper (as well as Refs.~\cite{Inomata:2016rbd, Kohri:2018awv,Inomata:2019zqy, Inomata:2019ivs}) and in Ref.~\cite{DeLuca:2019ufz}.  See the last part of Appendix D of Ref.~\cite{Espinosa:2018eve} for complementary information. (The notation of Ref.~\cite{DeLuca:2019ufz} is largely the same as that of Ref.~\cite{Espinosa:2018eve}.)

\begin{table}[htb!]  
\begin{center}
\caption{Comparison of notations. 
In this table, the wavenumber of the tensor and scalar perturbations  are denoted by $\bm{k}$ and $\bm{p}$, respectively, with $k = |\bm{k}|$ and $p = |\bm{p}|$.  $\bar{\eta}$ is the conformal time as an integration variable.
}
\begin{tabular}{| l | l | l | c |} 
\hline 
 This paper & Ref.~\cite{DeLuca:2019ufz} & correspondence  \\ \hline
 $v = p/k$  & $x = p/k$ & $v  \leftrightarrow x$ \\ \hline
  $u = |\bm{k}-\bm{p}|/k$ &   $y = |\bm{k}-\bm{p}|/k$ & $u  \leftrightarrow  y$  \\ \hline
 $x = k \eta$  & $z = k \eta /\sqrt{3}$ &  $x  \leftrightarrow  \sqrt{3} z$  \\ \hline
$\bar{x} = k \bar{\eta}$  & $u = k \bar{\eta} $ &  $\bar{x}  \leftrightarrow  u$  \\ \hline 
 $\mathcal{I} = x I = \mathcal{I}_\text{c} \cos x + \mathcal{I}_\text{s} \sin x$  & $\mathcal{I}_\text{c} \cos x + \mathcal{I}_\text{s} \sin x $ &  $ 9\, \mathcal{I}_\text{c} \leftrightarrow  \mathcal{I}_\text{c}$,  $ 9\,\mathcal{I}_\text{s} \leftrightarrow  \mathcal{I}_\text{s}$  \\ \hline
\end{tabular}
 \label{table:notation}
\end{center}
\end{table}

\section{Analytical expression of function $I(u,v,x)$ in TT gauge}\label{sec:analytic_formulae}
\subsection{Radiation-dominated era}
In a RD era, the Eq.~\eqref{I_definition} is integrated as follows:

\begin{align}
I (u,v,x) =& \frac{3}{ u^3 v^3 x^3} \left(  -6 u v x - u v \left(u^2 + v^2 -3 \right) x^2 \sin x  + 6 \sqrt{3} v \sin \frac{u x}{\sqrt{3}}  + 6 \sqrt{3} u \sin \frac{v x}{\sqrt{3}}  \right.  \nonumber \\
& - 6 \sqrt{3} u \cos \frac{u x}{\sqrt{3}} \sin \frac{v x}{\sqrt{3}} - 6 \sqrt{3} v \cos \frac{v x}{\sqrt{3}} \sin \frac{u x}{\sqrt{3}}  \nonumber \\
& \left.+ 3 (u^2 + v^2 -3) x \sin \frac{u x}{\sqrt{3}} \sin \frac{v x}{\sqrt{3}} + 6 u v x \cos \frac{u x}{\sqrt{3}} \cos \frac{v x}{\sqrt{3}} \right)  \nonumber \\
& + \frac{3(u^2 + v^2 -3)^2 }{ 4 u^3 v^3 x} \left(   \sin x  \left( -\text{Ci}\left( x+ \frac{(u+v)x}{\sqrt{3}} \right) -  \text{Ci}\left( \left| x- \frac{(u+v)x}{\sqrt{3}} \right| \right) \right. \right. \nonumber \\
& \left. + \text{Ci}\left( x+ \frac{(u-v)x}{\sqrt{3}} \right) +  \text{Ci}\left( x+ \frac{(-u+v)x}{\sqrt{3}} \right) + \log \left| \frac{3-(u+v)^2}{3-(u-v)^2} \right| \right)  \nonumber \\
&+ \cos x \left( \text{Si}\left( x+ \frac{(u+v)x}{\sqrt{3}} \right)   +  \text{Si}\left( x- \frac{(u+v)x}{\sqrt{3}} \right)  \right. \nonumber \\
& \left. -  \text{Si}\left( x+ \frac{(u-v)x}{\sqrt{3}} \right) -  \text{Si}\left( x+ \frac{(-u+v)x}{\sqrt{3}} \right)\right) ,
\end{align}
where $\text{Si}(x) \equiv \int_0^x \text{d} \bar{x} \frac{\sin \bar{x}}{\bar{x}}$ and $\text{Ci}(x) \equiv - \int_x^\infty \text{d} \bar{x} \frac{\cos \bar{x}}{\bar{x}}$.
The counterpart in the Newtonian gauge is Eq.~(22) of Ref.~\cite{Kohri:2018awv} [see also Ref.~\cite{Espinosa:2018eve}].

This formula can also be  expressed in the form
\begin{align}
x I (u,v,x) = \mathcal{I}_c (u,v,x) \cos x  + \mathcal{I}_s (u,v,x) \sin x,
\end{align}
where
\begin{align}
\mathcal{I}_\text{c} (u,v,x) =& \frac{1}{u^3 v^3 x^3} \left( 9 x \cos x \left(  -2 u v x +  2\sqrt{3} u \sin \frac{v x}{\sqrt{3}} + 2 \sqrt{3} v  \sin \frac{u x}{\sqrt{3}}  +  2 u v x \cos \frac{u x}{\sqrt{3}} \cos \frac{v x}{\sqrt{3}}    \right.   \right.  \nonumber  \\
& \left.     -2 \sqrt{3} u \cos \frac{u x}{\sqrt{3}} \sin \frac{v x}{\sqrt{3}}   -2 \sqrt{3} v \cos \frac{v x}{\sqrt{3}} \sin \frac{u x}{\sqrt{3}} +(u^2+v^2-3)x \sin \frac{u x}{\sqrt{3}} \sin \frac{v x}{\sqrt{3}}  \right)  \nonumber \\
&    - 3 \sin x \left(  6 u v x  \left( 1 + \cos \frac{u x}{\sqrt{3}}  +\cos \frac{v x}{\sqrt{3}}  \right) -12\sqrt{3} u \sin \frac{v x}{\sqrt{3}}  -12\sqrt{3} v \sin \frac{u x}{\sqrt{3}}  \right.  \nonumber \\
&   - 18 u v x \cos \frac{u x}{\sqrt{3}}  \cos \frac{v x}{\sqrt{3}}  + 3 (u^2 + v^2 +3 ) x \sin \frac{u x}{\sqrt{3}}  \sin \frac{v x}{\sqrt{3}} \nonumber \\
& \left. \left.  + \sqrt{3} u (12  + ( u^2 -  v^2 -3) x^2) \cos \frac{u x}{\sqrt{3}} \sin \frac{v x}{\sqrt{3}}  + (u \leftrightarrow v) \right) \right) \nonumber \\
& + \frac{3(u^2+v^2-3)^2 }{4u^3 v^3} \left( \text{Si}\left( x+ \frac{(u+v)x}{\sqrt{3}} \right)   +  \text{Si}\left( x- \frac{(u+v)x}{\sqrt{3}} \right)  \right. \nonumber \\
& \left. -  \text{Si}\left( x+ \frac{(u-v)x}{\sqrt{3}} \right) -  \text{Si}\left( x+ \frac{(-u+v)x}{\sqrt{3}} \right)\right) , \\
\mathcal{I}_\text{s} (u,v,x) =& \frac{1}{u^3 v^3 x^3} \left(  -3 uv (u^2 + v^2 -3 ) x^3  \right.  \nonumber \\  
&  + 9 x \sin x \left( -2 uv x  + 2\sqrt{3} v  \sin \frac{u x}{\sqrt{3}} +2\sqrt{3} u \sin \frac{v x}{\sqrt{3}}    -2\sqrt{3}u    \cos \frac{u x}{\sqrt{3}}  \sin \frac{v x}{\sqrt{3}} + (u \leftrightarrow v)   \right.  \nonumber \\
& \left.  + 2 uv x \cos \frac{u x}{\sqrt{3}} \cos \frac{v x}{\sqrt{3}}  + (u^2 + v^2 -3) x  \sin \frac{u x}{\sqrt{3}} \sin \frac{v x}{\sqrt{3}}   \right)  \nonumber \\
&  + 3 \cos x \left( 6 u v x \left( 1 +  \cos \frac{u x}{\sqrt{3}} + \cos \frac{v x}{\sqrt{3}}   \right)  -12\sqrt{3} v  \sin \frac{u x}{\sqrt{3}}     -12\sqrt{3} u  \sin \frac{v x}{\sqrt{3}} \right. \nonumber \\
&   -18 u v x  \cos \frac{u x}{\sqrt{3}}  \cos \frac{v x}{\sqrt{3}} + 3 (u^2 + v^2 +3) x  \sin \frac{u x}{\sqrt{3}}  \sin \frac{v x}{\sqrt{3}} \nonumber \\
& \left.\left.   +\sqrt{3} (12 + (u^2 - v^2 -3) x^2 )  \cos \frac{u x}{\sqrt{3}}  \sin \frac{v x}{\sqrt{3}}  + ( u \leftrightarrow v) \right) \right)     \nonumber \\ 
&- \frac{3(u^2+v^2-3)^2}{4u^3 v^3}\left( \text{Ci}\left( x+ \frac{(u+v)x}{\sqrt{3}} \right) +  \text{Ci}\left( \left | x- \frac{(u+v)x}{\sqrt{3}} \right | \right)  \right. \nonumber \\
& \left. -  \text{Ci}\left( x+ \frac{(u-v)x}{\sqrt{3}} \right) -  \text{Ci}\left( x+ \frac{(-u+v)x}{\sqrt{3}} \right) - \log \left| \frac{3-(u+v)^2}{3-(u-v)^2} \right| \right),
\end{align}
where  $(u \leftrightarrow v)$ denotes the term obtained by exchanging $u$ and $v$ in the previous term.

\subsection{Matter-dominated era}
\label{subsec:md_era_i_formula}

In the case of tensor perturbations induced in a MD era, we take the following normalizations for the transfer functions of $\psi$ and $\sigma$:
\begin{align}
\label{eq:psi_zeta_rel_md}
\psi (\eta, \bm{k}) = & -\frac{3}{5} \zeta (\bm{k}) T_{\psi} (\eta , k) , \\
\label{eq:sigma_zeta_rel_md}
\sigma (\eta, \bm{k} ) =& -\frac{3}{5} \zeta (\bm{k}) \frac{\sqrt{3}}{k} T_{\sigma} (\eta, k).
\end{align}
Then, the concrete expressions of the transfer functions are given as~\cite{DeLuca:2019ufz} 
\begin{align}
T_\psi (x) = &\frac{5}{3},  &  
T_\sigma (x)=& \frac{- x }{3 \sqrt{3}}.
\end{align}
Taking into account the relation $\Psi = - (3+3w)/(5+3w) \zeta$ on the superhorizon scales where $w= P/\rho$ is the equation-of-state parameter with $P$ denoting the pressure, we define $f(u,v,x)$ in a MD era $(3/5)^2 (2/3)^{-2} (=81/100)$ times Eq.~\eqref{fTT}. Strictly speaking, the coefficient of the last term should also be modified.  However, $T'_\psi$ vanishes in a MD era, so it does not matter. 
The explicit form of $f$ is $f(u,v,x) =  1- x^2/50 $, which grows in time. 

The function $I(u,v,x)$ in the MD era is
\begin{align}
I(u,v,x) = \frac{ - x^5 + 60 x^3 + 180 x \cos x  - 180 \sin x }{50 x^3}.
\end{align}
The late-time limit is $ I(u,v, x \to \infty ) = - x^2 /50$.  
Note that the counterpart in the Newtonian gauge is $6/5$ [Eq.~(36) of Ref.~\cite{Kohri:2018awv}; see also Ref.~\cite{Assadullahi:2009nf}].

\bibliographystyle{utphys}
\bibliography{gw_TT.bib}

\providecommand{\href}[2]{#2}\begingroup\raggedright\begin{thebibliography}{10}

\bibitem{Aghanim:2018eyx}
{\bfseries Planck} Collaboration, N.~Aghanim {\em et~al.}, ``{Planck 2018
  results. VI. Cosmological parameters},''
\href{http://arxiv.org/abs/1807.06209}{{\ttfamily arXiv:1807.06209
  [astro-ph.CO]}}.

\bibitem{Akrami:2018odb}
{\bfseries Planck} Collaboration, Y.~Akrami {\em et~al.}, ``{Planck 2018
  results. X. Constraints on inflation},''
\href{http://arxiv.org/abs/1807.06211}{{\ttfamily arXiv:1807.06211
  [astro-ph.CO]}}.

\bibitem{Alabidi:2012ex}
L.~Alabidi, K.~Kohri, M.~Sasaki, and Y.~Sendouda, ``{Observable Spectra of
  Induced Gravitational Waves from Inflation},''
  \href{http://dx.doi.org/10.1088/1475-7516/2012/09/017}{{\em JCAP} {\bfseries
  1209} (2012) 017},
\href{http://arxiv.org/abs/1203.4663}{{\ttfamily arXiv:1203.4663
  [astro-ph.CO]}}.

\bibitem{Alabidi:2013wtp}
L.~Alabidi, K.~Kohri, M.~Sasaki, and Y.~Sendouda, ``{Observable induced
  gravitational waves from an early matter phase},''
  \href{http://dx.doi.org/10.1088/1475-7516/2013/05/033}{{\em JCAP} {\bfseries
  1305} (2013) 033},
\href{http://arxiv.org/abs/1303.4519}{{\ttfamily arXiv:1303.4519
  [astro-ph.CO]}}.

\bibitem{Orlofsky:2016vbd}
N.~Orlofsky, A.~Pierce, and J.~D. Wells, ``{Inflationary theory and pulsar
  timing investigations of primordial black holes and gravitational waves},''
  \href{http://dx.doi.org/10.1103/PhysRevD.95.063518}{{\em Phys. Rev.}
  {\bfseries D95} no.~6, (2017) 063518},
\href{http://arxiv.org/abs/1612.05279}{{\ttfamily arXiv:1612.05279
  [astro-ph.CO]}}.

\bibitem{Inomata:2018epa}
K.~Inomata and T.~Nakama, ``{Gravitational waves induced by scalar
  perturbations as probes of the small-scale primordial spectrum},''
  \href{http://dx.doi.org/10.1103/PhysRevD.99.043511}{{\em Phys. Rev.}
  {\bfseries D99} no.~4, (2019) 043511},
\href{http://arxiv.org/abs/1812.00674}{{\ttfamily arXiv:1812.00674
  [astro-ph.CO]}}.

\bibitem{Byrnes:2018txb}
C.~T. Byrnes, P.~S. Cole, and S.~P. Patil, ``{Steepest growth of the power
  spectrum and primordial black holes},''
  \href{http://dx.doi.org/10.1088/1475-7516/2019/06/028}{{\em JCAP} {\bfseries
  1906} (2019) 028},
\href{http://arxiv.org/abs/1811.11158}{{\ttfamily arXiv:1811.11158
  [astro-ph.CO]}}.

\bibitem{Ben-Dayan:2019gll}
I.~Ben-Dayan, B.~Keating, D.~Leon, and I.~Wolfson, ``{Constraints on scalar and
  tensor spectra from $N_{eff}$},''
  \href{http://dx.doi.org/10.1088/1475-7516/2019/06/007}{{\em JCAP} {\bfseries
  1906} no.~06, (2019) 007},
\href{http://arxiv.org/abs/1903.11843}{{\ttfamily arXiv:1903.11843
  [astro-ph.CO]}}.

\bibitem{Mollerach:2003nq}
S.~Mollerach, D.~Harari, and S.~Matarrese, ``{CMB polarization from secondary
  vector and tensor modes},''
  \href{http://dx.doi.org/10.1103/PhysRevD.69.063002}{{\em Phys. Rev.}
  {\bfseries D69} (2004) 063002},
\href{http://arxiv.org/abs/astro-ph/0310711}{{\ttfamily arXiv:astro-ph/0310711
  [astro-ph]}}.

\bibitem{Ananda:2006af}
K.~N. Ananda, C.~Clarkson, and D.~Wands, ``{The Cosmological gravitational wave
  background from primordial density perturbations},''
  \href{http://dx.doi.org/10.1103/PhysRevD.75.123518}{{\em Phys. Rev.}
  {\bfseries D75} (2007) 123518},
\href{http://arxiv.org/abs/gr-qc/0612013}{{\ttfamily arXiv:gr-qc/0612013
  [gr-qc]}}.

\bibitem{Baumann:2007zm}
D.~Baumann, P.~J. Steinhardt, K.~Takahashi, and K.~Ichiki, ``{Gravitational
  Wave Spectrum Induced by Primordial Scalar Perturbations},''
  \href{http://dx.doi.org/10.1103/PhysRevD.76.084019}{{\em Phys. Rev.}
  {\bfseries D76} (2007) 084019},
\href{http://arxiv.org/abs/hep-th/0703290}{{\ttfamily arXiv:hep-th/0703290
  [hep-th]}}.

\bibitem{Assadullahi:2009nf}
H.~Assadullahi and D.~Wands, ``{Gravitational waves from an early matter
  era},'' \href{http://dx.doi.org/10.1103/PhysRevD.79.083511}{{\em Phys. Rev.}
  {\bfseries D79} (2009) 083511},
\href{http://arxiv.org/abs/0901.0989}{{\ttfamily arXiv:0901.0989
  [astro-ph.CO]}}.

\bibitem{Espinosa:2018eve}
J.~R. Espinosa, D.~Racco, and A.~Riotto, ``{A Cosmological Signature of the SM
  Higgs Instability: Gravitational Waves},''
  \href{http://dx.doi.org/10.1088/1475-7516/2018/09/012}{{\em JCAP} {\bfseries
  1809} no.~09, (2018) 012},
\href{http://arxiv.org/abs/1804.07732}{{\ttfamily arXiv:1804.07732 [hep-ph]}}.

\bibitem{Kohri:2018awv}
K.~Kohri and T.~Terada, ``{Semianalytic calculation of gravitational wave
  spectrum nonlinearly induced from primordial curvature perturbations},''
  \href{http://dx.doi.org/10.1103/PhysRevD.97.123532}{{\em Phys. Rev.}
  {\bfseries D97} no.~12, (2018) 123532},
\href{http://arxiv.org/abs/1804.08577}{{\ttfamily arXiv:1804.08577 [gr-qc]}}.

\bibitem{Cai:2018dig}
R.-g. Cai, S.~Pi, and M.~Sasaki, ``{Gravitational Waves Induced by non-Gaussian
  Scalar Perturbations},''
  \href{http://dx.doi.org/10.1103/PhysRevLett.122.201101}{{\em Phys. Rev.
  Lett.} {\bfseries 122} no.~20, (2019) 201101},
\href{http://arxiv.org/abs/1810.11000}{{\ttfamily arXiv:1810.11000
  [astro-ph.CO]}}.

\bibitem{Bartolo:2018rku}
N.~Bartolo, V.~De~Luca, G.~Franciolini, M.~Peloso, D.~Racco, and A.~Riotto,
  ``{Testing primordial black holes as dark matter with LISA},''
  \href{http://dx.doi.org/10.1103/PhysRevD.99.103521}{{\em Phys. Rev.}
  {\bfseries D99} no.~10, (2019) 103521},
\href{http://arxiv.org/abs/1810.12224}{{\ttfamily arXiv:1810.12224
  [astro-ph.CO]}}.

\bibitem{Unal:2018yaa}
C.~Unal, ``{Imprints of Primordial Non-Gaussianity on Gravitational Wave
  Spectrum},'' \href{http://dx.doi.org/10.1103/PhysRevD.99.041301}{{\em Phys.
  Rev.} {\bfseries D99} no.~4, (2019) 041301},
\href{http://arxiv.org/abs/1811.09151}{{\ttfamily arXiv:1811.09151
  [astro-ph.CO]}}.

\bibitem{Inomata:2019zqy}
K.~Inomata, K.~Kohri, T.~Nakama, and T.~Terada, ``{Gravitational Waves Induced
  by Scalar Perturbations during a Gradual Transition from an Early Matter Era
  to the Radiation Era},''
  \href{http://dx.doi.org/10.1088/1475-7516/2019/10/071}{{\em JCAP} {\bfseries
  1910} no.~10, (2019) 071},
\href{http://arxiv.org/abs/1904.12878}{{\ttfamily arXiv:1904.12878
  [astro-ph.CO]}}.

\bibitem{Inomata:2019ivs}
K.~Inomata, K.~Kohri, T.~Nakama, and T.~Terada, ``{Enhancement of Gravitational
  Waves Induced by Scalar Perturbations due to a Sudden Transition from an
  Early Matter Era to the Radiation Era},''
  \href{http://dx.doi.org/10.1103/PhysRevD.100.043532}{{\em Phys. Rev.}
  {\bfseries D100} no.~4, (2019) 043532},
\href{http://arxiv.org/abs/1904.12879}{{\ttfamily arXiv:1904.12879
  [astro-ph.CO]}}.

\bibitem{Yuan:2019udt}
C.~Yuan, Z.-C. Chen, and Q.-G. Huang, ``{Probing primordial–black-hole dark
  matter with scalar induced gravitational waves},''
  \href{http://dx.doi.org/10.1103/PhysRevD.100.081301}{{\em Phys. Rev.}
  {\bfseries D100} no.~8, (2019) 081301},
\href{http://arxiv.org/abs/1906.11549}{{\ttfamily arXiv:1906.11549
  [astro-ph.CO]}}.

\bibitem{Bartolo:2019oiq}
N.~Bartolo, D.~Bertacca, S.~Matarrese, M.~Peloso, A.~Ricciardone, A.~Riotto,
  and G.~Tasinato, ``{Anisotropies and non-Gaussianity of the Cosmological
  Gravitational Wave Background},''
\href{http://arxiv.org/abs/1908.00527}{{\ttfamily arXiv:1908.00527
  [astro-ph.CO]}}.

\bibitem{Yuan:2019wwo}
C.~Yuan, Z.-C. Chen, and Q.-G. Huang, ``{Log-dependent slope of scalar induced
  gravitational waves in the infrared regions},''
\href{http://arxiv.org/abs/1910.09099}{{\ttfamily arXiv:1910.09099
  [astro-ph.CO]}}.

\bibitem{Saito:2008jc}
R.~Saito and J.~Yokoyama, ``{Gravitational wave background as a probe of the
  primordial black hole abundance},''
  \href{http://dx.doi.org/10.1103/PhysRevLett.102.161101,
  10.1103/PhysRevLett.107.069901}{{\em Phys. Rev. Lett.} {\bfseries 102} (2009)
  161101}, \href{http://arxiv.org/abs/0812.4339}{{\ttfamily arXiv:0812.4339
  [astro-ph]}}.
[Erratum: Phys. Rev. Lett.107,069901(2011)].

\bibitem{Saito:2009jt}
R.~Saito and J.~Yokoyama, ``{Gravitational-Wave Constraints on the Abundance of
  Primordial Black Holes},'' \href{http://dx.doi.org/10.1143/PTP.126.351,
  10.1143/PTP.123.867}{{\em Prog. Theor. Phys.} {\bfseries 123} (2010)
  867--886}, \href{http://arxiv.org/abs/0912.5317}{{\ttfamily arXiv:0912.5317
  [astro-ph.CO]}}.
[Erratum: Prog. Theor. Phys.126,351(2011)].

\bibitem{Bugaev:2009zh}
E.~Bugaev and P.~Klimai, ``{Induced gravitational wave background and
  primordial black holes},''
  \href{http://dx.doi.org/10.1103/PhysRevD.81.023517}{{\em Phys. Rev.}
  {\bfseries D81} (2010) 023517},
\href{http://arxiv.org/abs/0908.0664}{{\ttfamily arXiv:0908.0664
  [astro-ph.CO]}}.

\bibitem{Clesse:2018ogk}
S.~Clesse, J.~García-Bellido, and S.~Orani, ``{Detecting the Stochastic
  Gravitational Wave Background from Primordial Black Hole Formation},''
\href{http://arxiv.org/abs/1812.11011}{{\ttfamily arXiv:1812.11011
  [astro-ph.CO]}}.

\bibitem{Wang:2019kaf}
S.~Wang, T.~Terada, and K.~Kohri, ``{Prospective constraints on the primordial
  black hole abundance from the stochastic gravitational-wave backgrounds
  produced by coalescing events and curvature perturbations},''
  \href{http://dx.doi.org/10.1103/PhysRevD.99.103531}{{\em Phys. Rev.}
  {\bfseries D99} no.~10, (2019) 103531},
\href{http://arxiv.org/abs/1903.05924}{{\ttfamily arXiv:1903.05924
  [astro-ph.CO]}}.

\bibitem{Cai:2019elf}
R.-G. Cai, S.~Pi, S.-J. Wang, and X.-Y. Yang, ``{Pulsar Timing Array
  Constraints on the Induced Gravitational Waves},''
  \href{http://arxiv.org/abs/1907.06372}{{\ttfamily arXiv:1907.06372
  [astro-ph.CO]}}.
[JCAP1910,no.10,059(2019)].

\bibitem{Bartolo:2019zvb}
N.~Bartolo, D.~Bertacca, V.~De~Luca, G.~Franciolini, S.~Matarrese, M.~Peloso,
  A.~Ricciardone, A.~Riotto, and G.~Tasinato, ``{Gravitational Wave
  Anisotropies from Primordial Black Holes},''
\href{http://arxiv.org/abs/1909.12619}{{\ttfamily arXiv:1909.12619
  [astro-ph.CO]}}.

\bibitem{Chen:2019xse}
Z.-C. Chen, C.~Yuan, and Q.-G. Huang, ``{Pulsar Timing Array Constraints on
  Primordial Black Holes with NANOGrav 11-Year Data Set},''
\href{http://arxiv.org/abs/1910.12239}{{\ttfamily arXiv:1910.12239
  [astro-ph.CO]}}.

\bibitem{Hajkarim:2019nbx}
F.~Hajkarim and J.~Schaffner-Bielich, ``{Thermal History of the Early Universe
  and Primordial Gravitational Waves from Induced Scalar Perturbations},''
\href{http://arxiv.org/abs/1910.12357}{{\ttfamily arXiv:1910.12357 [hep-ph]}}.

\bibitem{LIGOScientific:2018mvr}
{\bfseries LIGO Scientific, Virgo} Collaboration, B.~P. Abbott {\em et~al.},
  ``{GWTC-1: A Gravitational-Wave Transient Catalog of Compact Binary Mergers
  Observed by LIGO and Virgo during the First and Second Observing Runs},''
  \href{http://dx.doi.org/10.1103/PhysRevX.9.031040}{{\em Phys. Rev.}
  {\bfseries X9} no.~3, (2019) 031040},
\href{http://arxiv.org/abs/1811.12907}{{\ttfamily arXiv:1811.12907
  [astro-ph.HE]}}.

\bibitem{Audley:2017drz}
{\bfseries LISA} Collaboration, P.~Amaro-Seoane {\em et~al.}, ``{Laser
  Interferometer Space Antenna},''
\href{http://arxiv.org/abs/1702.00786}{{\ttfamily arXiv:1702.00786
  [astro-ph.IM]}}.

\bibitem{Janssen:2014dka}
G.~Janssen {\em et~al.}, ``{Gravitational wave astronomy with the SKA},''
  \href{http://dx.doi.org/10.22323/1.215.0037}{{\em PoS} {\bfseries AASKA14}
  (2015) 037},
\href{http://arxiv.org/abs/1501.00127}{{\ttfamily arXiv:1501.00127
  [astro-ph.IM]}}.

\bibitem{Hwang:2017oxa}
J.-C. Hwang, D.~Jeong, and H.~Noh, ``{Gauge dependence of gravitational waves
  generated from scalar perturbations},''
  \href{http://dx.doi.org/10.3847/1538-4357/aa74be}{{\em Astrophys. J.}
  {\bfseries 842} no.~1, (2017) 46},
\href{http://arxiv.org/abs/1704.03500}{{\ttfamily arXiv:1704.03500
  [astro-ph.CO]}}.

\bibitem{Matarrese:1997ay}
S.~Matarrese, S.~Mollerach, and M.~Bruni, ``{Second order perturbations of the
  Einstein-de Sitter universe},''
  \href{http://dx.doi.org/10.1103/PhysRevD.58.043504}{{\em Phys. Rev.}
  {\bfseries D58} (1998) 043504},
\href{http://arxiv.org/abs/astro-ph/9707278}{{\ttfamily arXiv:astro-ph/9707278
  [astro-ph]}}.

\bibitem{Arroja:2009sh}
F.~Arroja, H.~Assadullahi, K.~Koyama, and D.~Wands, ``{Cosmological matching
  conditions for gravitational waves at second order},''
  \href{http://dx.doi.org/10.1103/PhysRevD.80.123526}{{\em Phys. Rev.}
  {\bfseries D80} (2009) 123526},
\href{http://arxiv.org/abs/0907.3618}{{\ttfamily arXiv:0907.3618
  [astro-ph.CO]}}.

\bibitem{Domenech:2017ems}
G.~Domènech and M.~Sasaki, ``{Hamiltonian approach to second order gauge
  invariant cosmological perturbations},''
  \href{http://dx.doi.org/10.1103/PhysRevD.97.023521}{{\em Phys. Rev.}
  {\bfseries D97} no.~2, (2018) 023521},
\href{http://arxiv.org/abs/1709.09804}{{\ttfamily arXiv:1709.09804 [gr-qc]}}.

\bibitem{Gong:2019mui}
J.-O. Gong, ``{Analytic integral solutions for induced gravitational waves},''
\href{http://arxiv.org/abs/1909.12708}{{\ttfamily arXiv:1909.12708 [gr-qc]}}.

\bibitem{Tomikawa:2019tvi}
K.~Tomikawa and T.~Kobayashi, ``{On the gauge dependence of gravitational waves
  generated at second order from scalar perturbations},''
\href{http://arxiv.org/abs/1910.01880}{{\ttfamily arXiv:1910.01880 [gr-qc]}}.

\bibitem{DeLuca:2019ufz}
V.~De~Luca, G.~Franciolini, A.~Kehagias, and A.~Riotto, ``{On the Gauge
  Invariance of Cosmological Gravitational Waves},''
\href{http://arxiv.org/abs/1911.09689}{{\ttfamily arXiv:1911.09689 [gr-qc]}}.

\bibitem{Inomata:2016rbd}
K.~Inomata, M.~Kawasaki, K.~Mukaida, Y.~Tada, and T.~T. Yanagida,
  ``{Inflationary primordial black holes for the LIGO gravitational wave events
  and pulsar timing array experiments},''
  \href{http://dx.doi.org/10.1103/PhysRevD.95.123510}{{\em Phys. Rev.}
  {\bfseries D95} no.~12, (2017) 123510},
\href{http://arxiv.org/abs/1611.06130}{{\ttfamily arXiv:1611.06130
  [astro-ph.CO]}}.

\bibitem{Maggiore:1999vm}
M.~Maggiore, ``{Gravitational wave experiments and early universe cosmology},''
  \href{http://dx.doi.org/10.1016/S0370-1573(99)00102-7}{{\em Phys. Rept.}
  {\bfseries 331} (2000) 283--367},
\href{http://arxiv.org/abs/gr-qc/9909001}{{\ttfamily arXiv:gr-qc/9909001
  [gr-qc]}}.

\bibitem{Ma:1995ey}
C.-P. Ma and E.~Bertschinger, ``{Cosmological perturbation theory in the
  synchronous and conformal Newtonian gauges},''
  \href{http://dx.doi.org/10.1086/176550}{{\em Astrophys. J.} {\bfseries 455}
  (1995) 7--25},
\href{http://arxiv.org/abs/astro-ph/9506072}{{\ttfamily arXiv:astro-ph/9506072
  [astro-ph]}}.

\bibitem{Silk:1967kq}
J.~Silk, ``{Cosmic black body radiation and galaxy formation},''
\href{http://dx.doi.org/10.1086/149449}{{\em Astrophys. J.} {\bfseries 151}
  (1968) 459--471}.

\bibitem{Jeong:2014gna}
D.~Jeong, J.~Pradler, J.~Chluba, and M.~Kamionkowski, ``{Silk damping at a
  redshift of a billion: a new limit on small-scale adiabatic perturbations},''
  \href{http://dx.doi.org/10.1103/PhysRevLett.113.061301}{{\em Phys. Rev.
  Lett.} {\bfseries 113} (2014) 061301},
\href{http://arxiv.org/abs/1403.3697}{{\ttfamily arXiv:1403.3697
  [astro-ph.CO]}}.

\bibitem{Yuan:2019fwv}
C.~Yuan, Z.-C. Chen, and Q.-G. Huang, ``{Scalar Induced Gravitational Waves in
  Different Gauges},''
\href{http://arxiv.org/abs/1912.00885}{{\ttfamily arXiv:1912.00885
  [astro-ph.CO]}}.

\bibitem{Pitrou:2013hga}
C.~Pitrou, X.~Roy, and O.~Umeh, ``{xPand: An algorithm for perturbing
  homogeneous cosmologies},''
  \href{http://dx.doi.org/10.1088/0264-9381/30/16/165002}{{\em Class. Quant.
  Grav.} {\bfseries 30} (2013) 165002},
\href{http://arxiv.org/abs/1302.6174}{{\ttfamily arXiv:1302.6174
  [astro-ph.CO]}}.

\end{thebibliography}\endgroup

\end{document}